\documentstyle[preprint,aps]{revtex}

\tightenlines

\preprint{hep-th/9807168}

\draft

\title{
Black Hole Graybody Factor and Black Hole Entropy
}
\author{ Kenji Suzuki }
\address{
\it Department of physics, Tokyo Institute of Technology \\
\it Oh-okayama, Meguro, Tokyo 152, Japan \\
\it ks@th.phys.titech.ac.jp
}

\begin{document}
\maketitle

\begin{abstract}
We have proposed the entropy formula of the black hole 
which is constructed by two intersecting D-branes 
with no momentum,
whose compactification radii are 
constrained by the surface gravities in ten-dimensions.
We interpret the entropy of the black hole 
as the statistical entropy 
of the effective string. 
We further study the behavior 
of the absorption cross-section 
of the black hole using our entropy formula.

\end{abstract}


\section{Introduction}
The Bekenstein-Hawking entropies of the BPS saturated 
black holes have been studied in various dimensions 
from the nonperturbative aspects 
of string theory~\cite{CY,CT,SV,CM,HS}.
The black holes are constructed 
by the intersecting D-branes~\cite{CT,Tsey,GKT}.
The entropies of the extremal or near extremal 
black holes are interpreted 
as the statistical entropies 
in term of the levels of the microscopic states.

For example, the entropy of the black hole 
constructed by the intersecting D1-brane and D5-brane 
in ten-dimensions 
is interpreted as the statistical entropy 
which counts the degeneracy 
of the BPS saturated D-brane bound states.
In this calculation,
the level of the states 
is identified with the momentum~\cite{SV}.
The entropy formula in the microscopic D-brane picture 
is obtained for the large level of the states.
Therefore, this formula is correct for the large momentum.

At the same time, the entropy of the black hole 
constructed by the intersecting D1-brane and D5-brane 
is also explained as that 
of the effective string 
living on the D5-brane~\cite{KT,KM}.
The gas of strings has a total mass, some momentum 
and a winding number along the compactified direction,
which are expressed in terms of the charges of the D-branes.
The level of states is derived 
from the mass spectrum with the momentum 
and the winding number.
Then, the level of states is finite 
even with no momentum.
The entropy formula in term of the microscopic states 
is obtained for the large level of string states.
Therefore, this formula is correct for the large level of states 
even with no momentum.

In addition, the graybody factors for the black holes 
have been studied 
in various dimensions~\cite{KT,KM,DM,DMW,MS,GKl,CL}.
These are calculated by using the wave functions of 
the Klein-Gordon equation.
These are proportional to the entropies
of the black holes.
In~\cite{KT,KM},
the graybody factor of the effective string  
living on the D5-brane 
is shown to be the same as that calculated 
by using the wave functions of 
the Klein-Gordon equation 
in four- and five-dimensions.

On the other hand, the Euler numbers of the black holes have been 
investigated in four- and ten-dimensions~\cite{HHR,GK,LP,KS}.
We have calculated the Euler numbers of the black holes 
constructed by the intersecting D-branes in ten-dimensions~\cite{KS}.
These black holes have the compactification radii.
In this calculation,
we have proposed that the compactification radii 
are constrained by the surface gravities in ten-dimensions.
It is necessary to avoid the singular effects 
of the horizon in the compactified directions.
We have found that 
the compactification radii are the inverses of 
the surface gravity in the compactified directions 
using the Gauss-Bonnet theorem.
Using this determination, we have obtained the integer valued 
Euler numbers which are 
calculated by using the ten-dimensional Gauss-Bonnet action.
The Euler number is the sum of the Betti numbers.
The Betti numbers are integers.
Therefore, the Euler numbers must be also integers 
because of their definitions.
We emphasize that for arbitrary compactification radii,
the Euler numbers are not integers.

Moreover, we have calculated semiclassically 
the entropies of the black holes 
which are constructed by two intersecting D-branes 
with no momentum.
The black holes have the compactification radii 
which are constrained by the surface gravities~\cite{KS}.
We have obtained that the entropies of the black holes 
are proportional to the product of the quantized charges,
\begin{eqnarray}
 S = \pi Q_1Q_2 \ , \label{ent}
\end{eqnarray}
where $Q_1,Q_2$ are the quantized charges of two D-branes.
We have shown that the entropies are invariant 
under the T-duality transformation.
In the BPS limit, the entropies of the black holes 
constructed by two intersecting D-branes are finite.
Therefore, the black holes with two charges are more important 
than the others, and  
it is necessary to study the behaviors of the black holes with 
two charges.
However, the interpretation of the entropies (\ref{ent})
as the statistical entropies in term of 
the microscopic states has been still unclear.

The purpose of this paper is to interpret 
the entropy (\ref{ent}) of the black hole 
which is constructed 
by two intersecting D-branes 
with no momentum,
as the statistical entropy in term of the microscopic states.
The black hole has the compactified radii 
which are constrained by the surface gravities 
in ten-dimensions~\cite{KS}.
In order to explain the entropy 
of the black hole with no momentum as the statistical entropy,
we consider the effective string 
as mentioned above~\cite{KT}.
The gas of strings has a total mass, some momentum,
and a winding number along the compactified direction,
which are expressed in terms of the charges of the D-branes,
and the Newton's constant.
We interpret the entropy of the black hole as that 
of the effective string.
The level of the states is derived from the mass spectrum 
of the effective string obtained by the Virasoro constraints.
We consider the black hole in ten-dimensions,
which has six compactified directions.
The level of states is written 
in term of the charges,
the BPS parameter,
and the four-dimensional Newton's constant $G_4$.
The four-dimensional Newton's constant depends on 
six compactification radii.
We have proposed that the compactification radii are constrained 
by the surface gravities~\cite{KS}.
The entropy of the gas of strings is written 
in term of the levels of the states.
We rewrite the entropy of the effective string 
with the compactification radii 
which are constrained by the surface gravities.

Consequently, the entropy of the black hole 
which is constructed by two intersecting 
D3-branes with no momentum,
and has the compactification radii 
which are constrained  
by the surface gravities 
is shown to be the same as the statistical entropy 
of the effective string,
written in term of the levels of the effective string states.
In the BPS limit, the entropy of the black hole 
is finite.
We emphasize that 
the entropy of the black hole 
which is constructed by two intersecting D-branes 
with no momentum 
vanishes 
in the BPS limit 
for arbitrary compactification radii.

We further study the graybody factor for the black hole
with two large charges in four-dimensions.
The graybody factor is proportional to the entropy 
of the black hole.
We have shown that the entropies of the black holes 
constructed by the intersecting D-branes are invariant 
under T-duality transformation~\cite{KS}.
Therefore, the graybody factors are also invariant 
under the T-duality transformation.

The organization of this paper is as follows.
In section 2, we review the way to calculate 
the absorption cross-sections of the black holes 
with four charges in four-dimensions,
using the wave functions of the Klein-Gordon equation.
In section 3, we review the way to calculate 
the entropy and the absorption cross-section 
of the effective string.
We find that the entropy of the effective string 
is the same as that of the black hole 
as calculated in section 2.
In section 4, we review our proposal 
that the compactification radii are constrained 
by the surface gravities.
We calculate the entropy of the black hole 
which is constructed by two intersecting D3-branes 
with no momentum.
In section 5, we consider the entropy of the effective string 
with two large charges.
We show that the entropy of the black hole 
which is calculated in section 4 
can be interpreted as that of the effective string.
We further discuss the graybody factor for the black hole.


\section{Black hole graybody factor}
In this section, we review the way to obtain 
the graybody factor for the black hole 
using the wave functions 
which are the solutions of the Klein-Gordon wave equation.
This calculation is discussed in~\cite{KT,KM,DM,DMW,MS,GKl,CL}.

We consider the non-extremal black hole 
in four-dimensions.
The black hole is obtained by dimensional reduction 
from the ten-dimensional black hole.
We consider the black hole in ten-dimensions 
which is constructed by the intersecting 
D1-brane and D5-brane with the Kaluza Klein monopole 
and the momentum.
The black hole has two $U(1)$ charges 
$r_2,r_3$, and the momentum charge $r_1$,
and the Kaluza-Klein charge $r_n$,
and the BPS parameter $r_0$,
where $r_0, r_1, r_n \ll r_2, r_3 $.
The metric of the four-dimensional black hole 
obtained by dimensional reduction 
from the ten-dimensional black hole 
is that 
\begin{eqnarray}
 ds^2 &=& -f^{-1/2}hdt^2 + f^{1/2}(h^{-1}dr^2+ r^2d\Omega^2) \ ,\\
&& h(r) = (1-\frac{r_0}{r}) \ , \nonumber \\ 
&& f(r) = (1+\frac{r_1}{r})(1+\frac{r_2}{r})
(1+\frac{r_3}{r})(1+\frac{r_n}{r}) \ , \nonumber 
\end{eqnarray}
where 
\begin{eqnarray*}
 r_1 = r_0 \sinh ^2 \sigma_1 \ , \quad 
 r_n = r_0 \sinh ^2 \sigma_n \ .
\end{eqnarray*}
$\sigma_1, \sigma_n$ are the parameters.
The entropy of the black hole 
in the semiclassical calculation is 
\begin{eqnarray}
 S = A/4G_4 
  =  4\pi \sqrt{r_2r_3}r_0 \cosh\sigma_1\cosh\sigma_n /4G_4 \ .
\end{eqnarray}

We can also obtain the black holes 
which have four charges 
in ten- or eleven-dimensions.
In ten-dimensions, they are constructed 
by four intersecting D3-branes,
or by T-duality transformation from them.
In eleven-dimensions, they are constructed by 
three intersecting M5-branes with the momentum, 
or four intersecting M2-branes.
These are also constructed  by two intersecting M5-branes 
with the Kaluza-Klein monopole and momentum,
or M2-brane and M5-brane.

The $l$-th partial wave equation of a massless scalar is 
\begin{eqnarray}
\frac{h}{r^2}(hr^2\frac{dR}{dr}) 
+ \bigg[ f\omega^2 - h\frac{l(l+1)}{r^2}\bigg] R = 0 \ .
\end{eqnarray}
We consider the inner region (region I),
$r \ll r_2, r_3 $.
We have the equation 
\begin{eqnarray}
 z\frac{d}{dz}z\frac{dR_I}{dz} 
+ [D + \frac{C}{(1-z)} + \frac{E}{(1-z)^2}]R_I = 0 \ ,
\label{eq}
\end{eqnarray}
where 
\begin{eqnarray}
z &\equiv& h(r) \ , \nonumber \\
D &=& \omega^2r_2r_3\sinh^2\sigma_1 \sinh^2\sigma_n \ , \nonumber \\
C &=& \omega^2r_2r_3(\sinh^2\sigma_1 + \sinh^2\sigma_n)
 + l(l+1) \ , \nonumber \\
E &=& \omega^2r_2r_3-l(l+1) \ . 
\end{eqnarray}
If we propose the solution of the equation as 
\begin{eqnarray}
R_I = z^{\alpha}(1-z)^{\beta}F(z) \ ,
\end{eqnarray}
then we obtain the relations as 
\begin{eqnarray}
E + \beta(\beta-1) = 0 \ , \quad 
\alpha^2 + D + C + E = 0 \ .
\end{eqnarray}
We choose the solution as  
\begin{eqnarray}
\alpha = -i\omega\sqrt{r_2r_3}\cosh\sigma_1\cosh\sigma_n \ ,
 \nonumber \\
2\beta = 1 - \sqrt{(2l+1)^2-4\omega^2r_2r_3} \ .
\end{eqnarray}
The equation for $F$ is  
\begin{eqnarray}
z(1-z)\frac{d^2F}{dz^2}
+[(2\alpha+1)(1-z) - 2\beta z]\frac{dF}{dz}
-[(\alpha+\beta)^2 + D]F = 0 \ .
\end{eqnarray}
The hypergeometric equation is 
\begin{eqnarray}
z(1-z)\frac{d^2F}{dz^2} 
+ [C-(1+A+B)z]\frac{dF}{dz} - ABF = 0 \ ,
\end{eqnarray}
and we denote the solution of this equation as $F(A,B,C;z)$. 
Therefore, we obtain the solution of (\ref{eq}) as 
\begin{eqnarray}
R_I = z^{\alpha}(1-z)^{\beta}F(\alpha + \beta + i\sqrt{D} \ ,
\alpha + \beta - i\sqrt{D};1+2\alpha;z) \ . 
\end{eqnarray}
Using the asymptotics of the hypergeometric functions for 
$z\rightarrow 1$, we find that 
\begin{eqnarray}
 R_I\rightarrow
\left (\frac{r_0}{r}\right )^{\beta} 
\frac{\Gamma (1+2\alpha) \Gamma(1-2\beta)}
     {\Gamma (1+ \alpha- \beta -i \sqrt D) 
     \Gamma (1+ \alpha- \beta +i \sqrt D) } \ ,
\end{eqnarray}
for large $r$, and $R_I \sim z^\alpha(1-z)^\beta$ for small $r$.

We next consider the outer region (region II)
as $r \gg r_0, r_1, r_n$.
The wave equation is 
\begin{eqnarray}
\rho^{-2}\frac{d}{d\rho}\rho^2\frac{dR}{d\rho}
+ [1 + \frac{\omega(r_2 + r_3)}{\rho} 
+ \frac{\omega^2(r_2 r_3)-l(l+1)}{\rho^2}]R = 0 \ ,
\end{eqnarray}
where $\rho = \omega r$ . 
The solution of this equation 
which matches the asymptotic form in region I is 
\begin{eqnarray}
 R_{II} &=& A\rho^{-\beta}e^{\pm i\rho}F(s+1 \pm ni,2s+2;\mp2i\rho) \ ,
 \\ 
 &&n= -\frac{\omega(r_2+r_3)}{2} \ , \quad 
 - s(s+1) = \omega^2(r_2 r_3)-l(l+1) \ , \nonumber 
\end{eqnarray}
where $F$ is the confluent hypergeometric function.
We obtain that 
$ R_{II} \sim \frac{A}{2\rho}e^{\pm i\rho}$ 
for large $\rho$ and $l=0$,
and $R_{II} \sim A\rho^{-\beta}$ for small $\rho$.

Matching $R_{II}$ to $R_{I}$ in the range $r_0 \ll r \ll r_2, r_3$ ,
we find that 
\begin{eqnarray}
A = {(\omega r_0)^{\beta}}
\frac{\Gamma(1+2\alpha)\Gamma(1-2\beta)}
{\Gamma(1+\alpha-\beta-i\sqrt{D})
 \Gamma(1+\alpha-\beta+i\sqrt{D})} \ .
\end{eqnarray}
The flux per unit solid angle is 
\begin{eqnarray} 
   {\cal F} =  \frac{1}{2i} (R^* h r^2 \partial_r R 
  - {\rm c.c.}) \ .
\end{eqnarray}
 The absorption probability is the ratio of the incoming flux at
the horizon to the incoming flux at infinity,
\begin{eqnarray}
   P= \frac{{\cal F}_{\rm h}}
{{\cal F}_\infty^{\rm incoming}} 
    = {4\omega^2}\sqrt{r_2r_3}r_0 
 \cosh\sigma_1\cosh\sigma_n |A|^{-2} \ .
\end{eqnarray}
The s-wave absorption cross-section is 
\begin{eqnarray}
 \sigma_{\rm abs} &=& \frac{\pi}{\omega^2}P_{l=0} \nonumber \\
  &=& 4\pi \sqrt{r_2r_3}r_0 
  \cosh\sigma_2 \cosh\sigma_n|A|^{-2} \nonumber \\
  &=& 4\pi\sqrt{r_2r_3}r_0
  \cosh\sigma_2 \cosh\sigma_n 
  \frac{\omega}{2(T_L+T_R)}
  \frac{\left ( e^{\frac{\omega}{T_H}} - 1\right )}
 {\left (e^{\frac{\omega}{2 T_L}} - 1\right )
  \left (e^{\frac{\omega}{2 T_R}} - 1 \right ) }  \ ,
\label{cr1}
\end{eqnarray}
where we define the left and right temperatures as 
\begin{eqnarray}
T_R= \frac{1}{4\pi \sqrt{r_2r_3} \cosh(\sigma_1+\sigma_n)} \ , 
\quad 
T_L= \frac{1}{4\pi \sqrt{r_2r_3} \cosh(\sigma_1-\sigma_n)} \ , 
\label{tem}
\end{eqnarray}
and the Hawking temperature is
\begin{eqnarray}
T_H &=& \frac{1}{4\pi \sqrt{r_2r_3} \cosh\sigma_1\cosh\sigma_n} \ , 
\\
\frac{2}{T_H} &=& \frac{1}{T_L}+\frac{1}{T_L} \ .\nonumber 
\end{eqnarray}
For small $\omega$, the absorption cross-section coincides  
the area of the horizon.


\section{effective string}

In the section, we review the way to calculate 
the graybody factor 
and the statistical entropy 
of the effective string 
living on the D5-brane.
This calculation is discussed in~\cite{KM}.
We find that the entropy of the effective string 
is the same as that of the black hole 
as calculated semiclassically in the previous section.

We consider the effective string 
living on the D5-brane.
The gas of strings has a total mass $m$,
and has some momentum number $n_p$,
and a winding number $n_w$,
along the compactified direction 
with the compactification radii $R$.
We explain the entropy of the black hole as that 
of the effective string,
which has an effective tension 
$T_{eff}=1/(2\pi\alpha'_{eff})$ and an effective central charge 
$c_{eff}= N_B + \frac{1}{2}N_f$.
$N_B$ and $N_f$ are the numbers of bosons and fermions.
The levels of states, $N_L$ and $N_R$, are calculated 
from the mass spectrum of the effective string 
obtained by the Virasoro constraints.

We first consider the neutral absorption probability 
of the effective string.
We assume that the mass levels are given by 
\begin{eqnarray}
 m^2=
\left (2 \pi R n_w T_{\rm eff}+\frac{n_p}{R}\right )^2 
+8\pi T_{\rm eff} N_R=
\left (2 \pi R n_w T_{\rm eff}-\frac{n_p}{R}\right )^2 
+8\pi T_{\rm eff} N_L \ ,
\end{eqnarray}
which is obtaind from Virasoro constraint.
Let the incoming scalar have energy $\omega$. Since it does not 
carry any charge,
the numbers $n_w, n_p$ are not altered 
by the absorption of the scalar.
Thus we have 
\begin{eqnarray}
2m\delta m= 8\pi T_{\rm eff} \delta N_R
= 8\pi T_{\rm eff} \delta N_L \ .
\end{eqnarray}
The scalar contributes one left oscillator $\alpha_i$ and 
one right oscillator $\tilde\alpha_j$, with 
\begin{eqnarray}
i=j=\delta N_R=\delta N_L=
 \frac{m}{4\pi T_{\rm eff}}\omega \ .
\end{eqnarray}
We average the absorption rate over all
initial string states of a given mass.
We find that the absorption cross-section,
which is the absorption rate minus the emission rate, is
\begin{eqnarray}
 \sigma_{\rm abs}
= G_4 \delta N_L
\frac{e^{\beta_L^* \delta N_L + \beta_R^* \delta N_R}-1}
{\left (e^{\beta_L^* \delta N_L} -1\right) 
\left (e^{\beta_R^* \delta N_R} -1\right )} \ ,
\label{cr2}
\end{eqnarray}
where
\begin{eqnarray}
 \beta^*_{R,L}
= \frac{\partial S}{\partial N_{R,L}} 
= \frac{\pi}{\sqrt{N_{R,L}} } \ ,
\end{eqnarray}
and $G_4$ is the four-dimensional Newton's constant.
$S$ is the entropy of the effective string as
\begin{eqnarray}
 S = 2\pi(\sqrt{N_L} + \sqrt{N_R}) \ ,
\end{eqnarray}
with $c=6$ .
For the right movers, the factor in the exponent is 
\begin{eqnarray}
{\beta_R^*\delta N_R }=
 \frac{m\omega}{4 T_{\rm eff}\sqrt{N_R}} \ .
\end{eqnarray}
We identify this with $\frac{\omega}{2 T_R} $ ,
then 
\begin{eqnarray}
 T_R= 2 T_{\rm eff} \frac{\sqrt{N_R}}{m} \ .
\end{eqnarray}
We will identify the effective string 
tension as the D-string tension, as 
\begin{eqnarray}
T_{\rm eff} =\frac{1}{8\pi r_2 r_3} \ ,
\end{eqnarray}
where we have set $\alpha'=1$.
We obtain that 
\begin{eqnarray}
T_R &=& \frac{1}{4\pi \sqrt{r_2r_3} \cosh(\sigma_1+\sigma_n)} \ ,
\nonumber \\ 
T_L &=& \frac{1}{4\pi \sqrt{r_2r_3}\cosh(\sigma_1-\sigma_n)} \ ,
\end{eqnarray}
by using 
\begin{eqnarray}
 m &=& \frac{r_0}{8G_4}
 (\cosh(2\sigma_1) + \cosh(2\sigma_n)) \ , \nonumber \\
 2\pi R n_{\omega}T_{eff} 
   &=& \frac{r_0}{8G_4} \sinh(2\sigma_1) \ , \nonumber \\ 
 \frac{n_p}{R}
   &=& \frac{r_0}{8G_4} \sinh(2\sigma_n) \ . \label{m}
\end{eqnarray}
These are in agreement 
with the temperatures (\ref{tem}) as discussed in previous section.
The cross-section (\ref{cr2}) is in complete 
agreement with the result (\ref{cr1}).
The entropy is 
\begin{eqnarray}
 S = 2\pi \sqrt{\frac{c_{eff}}{6}}(\sqrt{N_R} + \sqrt{N_L})
   = 4\pi \sqrt{r_2r_3}r_0 \cosh\sigma_1\cosh\sigma_n /4G_4 \ ,
\end{eqnarray}
with $c_{eff} = 6$, which is in agreement with the entropy
caluculated semiclassically in the previous section.


We next consider the absorption probability 
of the charged particles.
It was found that the wave equation for
a scalar of energy $k_0$ and charge $k_5$ is obtained from that 
for a neutral scalar of energy $\omega$ by the following
replacement of parameters:
\begin{eqnarray}
\omega \rightarrow \omega' , \qquad \sigma_n \rightarrow
\sigma'_n ,
\end{eqnarray}
where 
\begin{eqnarray}
 \omega'=\sqrt{ k_0^2 - k_5^2} \ , \qquad
e^{\sigma'_n} = e^{\sigma_n} \frac{k_0- k_5}{\omega'} \ .
\end{eqnarray}
The incoming scalar carries the $U(1)$ charge corresponding
to the momentum component $k_5$, which is
along the direction in which the effective string
carries the momentum and the winding number.
The variation $\delta n_p$, whose relation
to the absorbed charge is
$ k_5 = \frac{\delta n_p}{R} $ ,
while $ k_0= \delta m$ .
Thus, we find that 
\begin{eqnarray}
2m k_0= 8\pi T_{\rm eff} \delta N_R+
2 k_5 \left (2 \pi R n_w T_{\rm eff}+\frac{n_p}{R}\right ) \ ,
\nonumber \\
2m k_0= 8\pi T_{\rm eff} \delta N_L
-2 k_5 \left (2 \pi R n_w T_{\rm eff}-\frac{n_p}{R}\right ) \ .
\end{eqnarray}
Using this, we obtain that 
\begin{eqnarray}
 \beta_L^*\delta N_L 
= 2\pi \sqrt{r_2r_3} \left ( k_0 \cosh (\sigma_1 -\sigma_n)
+ k_5 \sinh (\sigma_1 -\sigma_n)\right )
= \frac{\omega'}{2 T'_L} \ ,
\nonumber \\
 \beta_R^*\delta N_R 
= 2\pi \sqrt{r_2r_3} \left ( k_0 \cosh (\sigma_1 +\sigma_n)
- k_5 \sinh (\sigma_1 +\sigma_n)\right ) 
=\frac{\omega'}{2 T'_R} \ ,
\end{eqnarray}
where 
\begin{eqnarray}
\frac{1}{T'_L} &=& 4\pi \sqrt{r_2r_3} \cosh (\sigma_1 - \sigma'_n) \ ,
\quad 
\frac{1}{T'_R}= 4\pi \sqrt{r_2r_3} \cosh (\sigma_1+ \sigma'_n) \ , \\
\frac{1}{T'_H} &=& \frac{1}{2 T'_L}+\frac{1}{2 T'_R} \nonumber \ .
\end{eqnarray}
We then find that the effective string 
absorption cross-section for charged 
scalars can be written as 
\begin{eqnarray}
 \sigma_{\rm abs} = 4\pi \sqrt{r_2r_3} r_0 
\cosh \sigma_1\cosh \sigma'_n 
  \frac{\omega'}{2(T'_L+T'_R)}
  \frac{\left ( e^{\frac{\omega'}{T'_H}} - 1\right )}
 {\left (e^{\frac{\omega'}{2 T'_L}} - 1\right )
  \left (e^{\frac{\omega'}{2 T'_R}} - 1 \right ) } \ . 
\end{eqnarray}
The result is in agreement 
with that calculated by using the wave functions 
of the Klein-Gordon wave equation.


\section{surface gravity and compactification}
In this section, we review the way 
to define the compactification radii 
by the surface gravity, as discussed in~\cite{KS}.
We further calculate semiclassically the entropy 
of the black hole 
constructed by two intersecting D-branes with no momentum.

We have proposed the determination of radii 
in the compactified directions 
by the surface gravity in ten-dimensions.
It is necessary to avoid the singular effects 
of the horizon in the compactified directions.
We recall the way 
to define the period in the time direction.
If no singular effects of the horizon exist in $t-r$ directions,
then the topology of these directions is ${\bf R^2}$,
and the Euler number of these directions is 1.
Using the Gauss-Bonnet theorem,
we have found that 
the Euclidean time coordinate 
have the period which is 
the inverse of the surface gravity.
Therefore we need to take this period 
in the time direction 
to avoid the singular effects of the horizon.
Similarly, we need to fix 
the compactification radii to avoid the singular effects 
of the horizon in the compactified directions.
We have found that 
the compactification radii are the inverses of 
the surface gravity in the compactified directions 
using the Gauss-Bonnet theorem.
Using the determination of radii,
we have obtained the integer valued 
Euler numbers which are calculated 
by using the ten-dimensional Gauss-Bonnet action.
The Euler number is the sum of the Betti numbers.
The Betti numbers are integers.
Therefore, the Euler numbers must be also integers 
because of their definition.
We emphasize that for arbitrary compactification radii,
the Euler numbers are not integers.

We have proposed that the compactification radius 
$\beta_i$ in the $i$-th direction in ten-dimensions 
is constrained 
by the following "surface gravity $\kappa_i$":
\begin{eqnarray}
  \beta_i(r_H) &=& \frac{2\pi}{\kappa_i} \ , \quad 
  \kappa_{i}(r_H)
   \equiv \frac{1}{2}\frac{\partial _r g_{ii}}
    {\sqrt{g_{rr}g_{ii}}}\bigg|_{r = r_H} \ , \nonumber \\
  && (i=1 \cdots 6) \ .
\end{eqnarray}
We define the proper length in the $i$-th direction as:
\begin{eqnarray}
  L_i(r) &\equiv& 
\bigg|\int_{0}^{\beta_{i}(r)}\sqrt{g_{ii}}dx_i \bigg| 
 = 4\pi \bigg| \frac{\sqrt{g_{rr}}}{\partial_r (\ln g_{ii})} 
\bigg| \ , \nonumber \\
  && ( i= 1,\cdots, 6 ) \ .
\end{eqnarray}
We consider the black hole constructed 
by the two intersecting D3-branes in ten-dimensions 
with the following metric,
\begin{eqnarray}
   ds^2_{10}  &=& (H_1 H_2)^{1/2} 
    \big[ H_1^{-1} H_2^{-1} (- hdt^2 +  dx_1^2 ) + dx_2^2 
  \nonumber \\
  &&  \quad + H_1^{-1} (dx_3^2 + dx_4^2)
       + H_2^{-1} (dx_5^2 + dx_6^2) 
       + h^{-1} dr^2 + r^2 d\Omega^2_2 \big] \ , \nonumber \\
 &&  h = (1-\frac{r_0}{r}) \ ,\quad 
     H_i = 1+ \frac{r_i}{r} \ , \quad (i=1,2) 
\end{eqnarray}
where $r_0 \ll r_i$. 
The four-dimensional black hole 
with two large charges is obtained by dimensional reduction 
from the ten-dimensional black hole.
These proper lengths are 
\begin{eqnarray}
 L_3 = L_4 = L_5 = L_6 
 &=& 8\pi(r + r_1)^{5/4} (r + r_2)^{5/4}/(r_1-r_2)(r-r_0)^{1/2} \ ,
  \nonumber \\
 L_1 = L_2 
 &=& 8 \pi (r + r_1)^{5/4} (r + r_2)^{5/4} 
 r /(r-r_0)^{1/2}[r_1(r+r_2) + r_2(r+r_1)] . \label{L}
\end{eqnarray}
The area in the $\theta-\phi$ directions is 
\begin{eqnarray}
 A_{\theta\phi}
 = \int_{0}^{\pi} d\theta \int_{0}^{2\pi} d\phi  
   \sqrt{g_{\theta\theta}g_{\phi\phi}}
 = 4 \pi (r + r_1)^{1/2} (r + r_2)^{1/2}r \ .
\end{eqnarray}
Then we find that the entropy of two intersecting D-branes 
is given by 
\begin{eqnarray}
S &=& A_8/4G_{10} \bigg|_{r=r_0}
   = L_1L_2L_3L_4L_5L_6 A_{\theta\phi}/4G_{10} \bigg|_{r=r_0} 
  \nonumber \\
  &=& (8\pi)^6\pi (r+r_1)^8(r+r_2)^8 r^3\nonumber \\
  && /G_{10}(r-r_0)^3 \bigg[(r_1-r_2)^4[r(r_1+r_2)+2r_1r_2]^2 
  \bigg] \bigg|_{r=r_0}\ ,\label{DD}
\end{eqnarray}
where $G_{10}$ is the ten-dimensional Newton's constant.
We consider the entropy for the extremal black holes,
namely, $h=1$ and $r_0 = 0$.
We first rewrite $(r-r_0)^3 \to r^3$ in the equation 
(\ref{DD}), and we next insert $r = r_0 = 0$.
We obtain the finite entropy for the extremal black holes as 
\begin{eqnarray}
S &=& 4(4\pi)^7 r_1^6r_2^6 
 /(r_1-r_2)^4 ,
\label{ex} 
\end{eqnarray}
with $G_{10}= 1$ .
We define the quantized charges of Dp-branes as
\begin{eqnarray}
 Q_i  = r_i/c_p, \label{cha}
\end{eqnarray}
where $c_p$ is 
\begin{eqnarray}
&& c_p = G_4 M_p |_{r \to 0}, \quad 
   M_p = \beta_{i1} \cdots \beta_{ip} , \label{Mp} \\
&& G_4 = G_{10}/(L_1L_2L_3L_4L_5L_6). \nonumber 
\end{eqnarray}
$i_1, \cdots i_p$ are the directions which the D-brane wrapping on.
$M_p$ is the bare mass of Dp-brane.
Then the length in the definition of the mass (\ref{Mp})
is not the proper length $L_i$, but the bare length $\beta_i$.
On the other hand, $G_{4}$ is given by 
$ \frac{1}{G_{4}}  
= \frac{1}{G_{10}} \int_0^{\beta_1} dx_1 \cdots \int_0^{\beta_6} dx_6 
 \sqrt{g_{11} \cdots g_{66}},$
which is obtained by the dimensional reduction of the action.
Then the quantized charges $Q_1$ and $Q_2$ for D3-branes are 
\begin{eqnarray}
 &&Q_1 = L_1L_2L_3L_4L_5L_6/\beta_1\beta_3\beta_4 r_1 |_{r=0}\ 
       = (8\pi)^2 4\pi(r_1r_2)^{3}/(r_1-r_2)^2 , 
\nonumber \\ 
 &&Q_2 = L_1L_2L_3L_4L_5L_6/\beta_1\beta_5\beta_6 r_2 |_{r=0}
       = (8\pi)^2 4\pi(r_1r_2)^{3}/(r_1-r_2)^2 \ ,
\end{eqnarray}
with $G_{10} = 1$. 
Using these charges, we rewritten the entropy (\ref{ex})
\begin{eqnarray}
S= \pi Q_1Q_2  .
\end{eqnarray}
We also obtain the finite quantized charges for two D4-branes 
and for the two D2-branes using the definition (\ref{cha}).
For two D4-branes, the metric and the quantized charges are 
\begin{eqnarray}
   ds^2_{10}  &=& (H_1 H_2)^{1/2} 
    \big[ H_1^{-1} H_2^{-1} (- dt^2 +  dx_1^2  + dx_2^2) 
  \nonumber \\
  &&  \quad + H_1^{-1} (dx_3^2 + dx_4^2)
       + H_2^{-1} (dx_5^2 + dx_6^2) 
       + dr^2 + r^2 d\Omega^2_2 \big] \ ,  \\
 &&   H_i = 1+ \frac{r_i}{r} \ , \quad (i=1,2) \nonumber \\
 Q_1 
        &=& (8\pi)^2 (r_1r_2)^{5/2}/(r_1-r_2)^2 , \nonumber \\ 
 Q_2  
      &=& (8\pi)^2 (r_1r_2)^{5/2}/(r_1-r_2)^2 ,
\end{eqnarray}
and the entropy is rewritten as 
\begin{eqnarray}
 S = \pi Q_1 Q_2 / T^2,
\end{eqnarray}
where $T$ is the temperature as $T = \frac{1}{4\pi \sqrt{r_1r_2}}$.
For two D2-branes, the metric and the quantized charges are 
\begin{eqnarray}
   ds^2_{10}  &=& (H_1 H_2)^{1/2} 
    \big[ H_1^{-1} H_2^{-1} (- dt^2 )+  dx_1^2  + dx_2^2 
  \nonumber \\
  &&  \quad + H_1^{-1} (dx_3^2 + dx_4^2)
       + H_2^{-1} (dx_5^2 + dx_6^2) 
       + dr^2 + r^2 d\Omega^2_2 \big] \ ,  \\
 &&   H_i = 1+ \frac{r_i}{r} \ , \quad (i=1,2) \nonumber \\
 Q_1 
       &=& (8\pi)^2 (4\pi)^2(r_1r_2)^{7/2}/(r_1-r_2)^2 , 
\nonumber \\ 
 Q_2 
       &=& (8\pi)^2 (4\pi)^2(r_1r_2)^{7/2}/(r_1-r_2)^2 ,
\end{eqnarray}
and the entropy is rewritten as 
\begin{eqnarray}
 S = \pi Q_1 Q_2 T^2. 
\end{eqnarray}
In three cases, the charges $Q_1$,$Q_2$ are constrained to be 
same, respectively.
The proper lengths generally satisfy the conditions as 
$L_1 = L_2$, and $L_3 = L_4 = L_5 = L_6$,
if the entropy is invariant under the T-dualty transformation.
Therefore the ratios of charges for these branes are 
\begin{eqnarray}
 \frac{Q_1}{Q_2} = \frac{r_1}{r_2} 
  \sqrt{\frac{g_{33}g_{44}}{g_{55}g_{66}}} \bigg|_{r = 0} = 1,
\end{eqnarray}
using the definition (\ref{Mp}).
Then this constraint is common 
for the quantized charges of these branes.
Therefore we obtain the new constraint 
for the quantized charges.



\section{Black hole entropy with two large charges}

In this section,
we show that the entropy of the black hole 
which is constructed by two intersecting D-branes 
with no momentum,
is the same as the statistical entropy 
of the effective string 
as discussed in section 3.
The black hole has the compactification radii 
which are constrained by the surface gravities.

We consider the black hole constructed by two D3-branes. 
These have the finite values of quantized charges in this section.
In section 3, we consider the black hole constructed 
by the D1-brane and the D5-brane.
However, the quantized charges for the branes are not finite 
using the compactification radii as discussed in section 4.
Further, we consider the effective strings 
for the black hole constructed by two D3-branes.
These are living on the intersecting direction of two D-branes.
The result of calculations for the black hole is same as 
the discussed in section 3.

The entropy of the effectve strings 
is written in term of the levels of states.
The levels are derived from the mass spectrum,
which are written by the two charges $r_2,r_3$,
the BPS parameter $r_0$,
and the four-dimensional Newton's constant $G_4$,
as discussed in section 3.
The four-dimensional Newton's constant depends on 
the compactification radii.
We have proposed that the compactification radii are constrained 
by the surface gravities~\cite{KS}.
We rewrite the entropy of the effective string 
with the compactification radii 
which are constrained by the surface gravities.
The entropy is shown to be the same 
with that of the black hole as calculated semiclassically 
in previous section.

In addition, we calculate the graybody factor 
for the black hole.
The graybody factor is proportional 
to the entropy of the black hole.
We have shown that the entropies of the black holes constructed 
by the intersecting D-branes are invariant 
under the T-duality transformation~\cite{KS}.
Therefore, the graybody factors for the black holes are 
also invariant under the T-duality transformation.

We consider the black hole with two large charges,
no Kaluza-Klein charge and no momentum,
which is reviewed in section 2, with $r_1, r_n = 0 \quad 
( \sigma_1 , \sigma_n = 0)$.
The temperatures are 
\begin{eqnarray}
T_L = T_R = \frac{1}{4\pi\sqrt{r_2r_3}} \ .
\end{eqnarray}
The levels of states are 
\begin{eqnarray}
\sqrt{cN_R/6} = \sqrt{cN_R/6}= (\frac{1}{8\pi T_{\rm eff}})^{1/2}m 
       = \sqrt{r_2r_3} 2 r_0 / 8G_4 ,
\end{eqnarray}
from (\ref{m}) with $\sigma_1,\sigma_n \to 0$ .
Therefore, the entropy of the effective string 
is shown to be the same as the entropy of the black hole 
as calculated semiclassically in four-dimensions, namely 
\begin{eqnarray}
 S = 2\pi(\sqrt{cN_R/6} + \sqrt{cN_L/6}) 
= 4\pi \sqrt{r_2r_3} r_0 /4G_4 \ ,
\end{eqnarray}
with $c=6$.
Using the proper lengths $L_i$, where $i= 1,\cdots 6$,
the four-dimensional Newton's constant is written as 
\begin{eqnarray}
G_4 = G_{10}/L_1L_2L_3L_4L_5L_6 \ .
\end{eqnarray}

We consider the entropy of the extremal black hole.
In the calculation of the proper length,
we consider the black hole constructed 
by the two intersecting D3-branes in ten-dimensions 
as the following metric,
\begin{eqnarray}
   ds^2_{10}  &=& (H_1 H_2)^{1/2} 
    \big[ H_1^{-1} H_2^{-1} (- dt^2 +  dx_1^2 ) + dx_2^2 
  \nonumber \\
  &&  \quad + H_1^{-1} ( dx_3^2 + dx_4^2 ) 
            + H_2^{-1} ( dx_5^2 + dx_6^2 ) 
       +   dr^2 + r^2 d\Omega^2_2 \big] \ ,
 \nonumber \\
  &&  H_i = 1+ \frac{r_i}{r} \ . \quad (i=1,2) 
\end{eqnarray}
as discussed in the previous section with $r_0 = 0$.
The four-dimensional black hole 
with two large charges is obtained by dimensional reduction 
from the ten-dimensional black hole.
We introduce the stretched horizon $ r = a $ 
in order to obtain the finite state levels.
We suppose that the event horizon and the mass are varied 
because of the quantum correction of the gravity.
Then the mass for the effective string $m$ is written as 
\begin{eqnarray}
 && m = \frac{r_2 + r_3 + \epsilon}{8G_4} - m_{bare}\ ,\\
 && m_{bare} = \frac{r_2 + r_3}{8G_4} \nonumber
\end{eqnarray}
where $m_{bare}$ is the bare mass, 
and the the term $\epsilon$ is the quantum effect for the bare mass. 
Then, the levels of the states are 
\begin{eqnarray}
\sqrt{N_L} &=& \sqrt{N_R}= (\frac{1}{8\pi T_{\rm eff}})^{1/2}m 
 \nonumber \\
      &=& 2\sqrt{r_1r_2} \epsilon / 8G_4 |_{r \to a}
 \nonumber \\
      &=& 2\sqrt{r_1r_2} \epsilon L_1L_2L_3L_4L_5L_6/ 8G_{10}
 \bigg|_{r \to a} \nonumber \\
      &=& 2(8\pi)^6 \epsilon \sqrt{r_1r_2}(r+r_1)^{15/2}(r+r_2)^{15/2}
\nonumber \\
  && /8G_{10} r \bigg[(r_1-r_2)^4[r(r_1+r_2)+2r_1r_2]^2
   \bigg]  \bigg|_{r \to a}\ ,
\end{eqnarray}
where the proper length $L_i$  
are the same as that in (\ref{L}) with $r_0 \to 0$.
In this calculation, we have used the fact that 
the compactification radii 
are constrained by the surface gravities.
The entropy is rewritten as 
\begin{eqnarray}
 S &=& 2\pi(\sqrt{cN_R/6} + \sqrt{cN_L/6}) 
    =  4\pi \sqrt{cr_2r_3/6} \epsilon / 4G_4 \bigg|_{r \to a}\  
        \nonumber \\
   &=& (8\pi)^6 \pi \sqrt{cr_1r_2/6} \epsilon 
       (r +r_1)^{15/2}(r+r_2)^{15/2}
\nonumber \\
  && /G_{10} r \bigg[(r_1-r_2)^4[r(r_1+r_2)+2r_1r_2]^2 
   \bigg]  \bigg|_{r \to a}
 \label{ES}.
\end{eqnarray}
%
Then we obtain the finite entropy 
which is the same as the entropy (\ref{ex}) with 
$ a \sim \epsilon$ and $\epsilon \to 0$.
Therefore, the entropy of the black hole  
with two charges, no momentum,
and with the compactification radii 
which are constrained by the surface gravities, 
is shown to be the same 
as the statistical entropy of the effective string,
which is written in term of the levels 
of the effective string states.
In the BPS limit, the entropy of the black hole 
which we have just considered is finite.
We emphasize that for arbitrary compactification radii,
the entropies of the black holes with two charges and no momentum 
vanish in the BPS limit.

We next consider the absorption cross-section of 
the extremal black hole with two large charges.
The absorption cross-section in this range is 
\begin{eqnarray}
 \sigma_{\rm abs} &=& \frac{\pi}{\omega^2}P_{l=0} 
  = 4\pi\sqrt{r_2r_3} \epsilon \frac{\omega}{2(T_L+T_R)} 
  \frac{\left ( e^{\frac{\omega}{T_H}} - 1\right )}
 {\left (e^{\frac{\omega}{2 T_L}} - 1\right )
  \left (e^{\frac{\omega}{2 T_R}} - 1 \right ) } \nonumber \\
  &=& 4\pi \sqrt{r_2r_3} \epsilon G_4 \frac{\omega}{2(T_L+T_R)} 
        \frac{\left ( e^{\frac{\omega}{T_H}} - 1\right )}
        {\left (e^{\frac{\omega}{2 T_L}} - 1\right )
        \left (e^{\frac{\omega}{2 T_R}} - 1 \right ) } 
        \frac{L_1L_2L_3L_4L_5L_6}{G_{10}} \nonumber \\
  &=& 4SG_4 \frac{\omega}{2(T_L+T_R)} 
        \frac{\left ( e^{\frac{\omega}{T_H}} - 1\right )}
        {\left (e^{\frac{\omega}{2 T_L}} - 1\right )
        \left (e^{\frac{\omega}{2 T_R}} - 1 \right ) } \ ,
\end{eqnarray} 
with $T_H =T_L = T_R = \frac{1}{4\pi\sqrt{r_2r_3}}$.
%
We have shown that the entropy of the black hole constructed by 
the intersecting D-branes is invariant 
under the T-duality transformation~\cite{KS}.
Therefore, the cross-section is also invariant
under the T-duality transformation.
In the extremal case, 
we obtain the cress-section for the black holes 
constructed by two D3-branes as 
\begin{eqnarray}
\sigma_{abs} 
 &=& 4SG_4 \frac{\omega}{2(T_L+T_R)} 
        \frac{\left ( e^{\frac{\omega}{T_H}} - 1\right )}
        {\left (e^{\frac{\omega}{2 T_L}} - 1\right )
        \left (e^{\frac{\omega}{2 T_R}} - 1 \right ) } 
\nonumber \\
   &=& 
        \frac{ 4\pi Q_1Q_2 G_4 \omega}{2(T_L+T_R)} 
        \frac{\left ( e^{\frac{\omega}{T_H}} - 1\right )}
        {\left (e^{\frac{\omega}{2 T_L}} - 1\right )
        \left (e^{\frac{\omega}{2 T_R}} - 1 \right ) } \ .
\end{eqnarray}
These are proportional to the products of the quantized charges.

\section{Conclusion}

We have proposed the entropy formula of the black hole (\ref{ent}).
The black hole is constructed 
by two intersecting D-branes 
with no momentum, 
and has the compactification radii 
which are constrained by the surface gravities in ten-dimensions.
We have interpreted the entropy of the black hole 
as the statistical entropy of the effective string. 
The gas of the strings has a total mass, and has some momentum 
and a winding numbers along the compactified direction.

First, we have considered the black hole with four charges 
in ten-dimensions.
The black hole is constructed 
by the intersecting D1-brane and D5-brane 
with momentum and the Kaluza-Klein monopole.
We have reviewed the way to obtain the graybody factor 
using the wave functions of the Klein-Gordon equation.
The graybody factor is shown to be propotional 
to the area of the black hole horizon.

Second, we have reviewed the way to calculate 
the graybody factor of the effective string 
living on the D5-brane~\cite{KT}.
The gas of strings has a total mass, some momentum,
and a winding number along the compactified direction.
We have explained the entropy using the effective string,
which has an effective tension 
$T_{eff}=1/(2\pi\alpha'_{eff})$ and an effective central charge 
$c_{eff}= N_B + \frac{1}{2}N_f$.
We have found that the graybody factor of the effective string  
is the same as that for the black hole by 
using the wave functions of the Klein-Gordon equation.

Next, we have considered the entropy of the black hole 
which is constructed 
by two intersecting D3-branes 
with no momentum and winding number.
We have proposed that 
the black hole has the compactification radii 
which are constrained by the surface gravities.
The entropy is proportional to the product 
of two quantized charges of D-branes.
In this paper,
we are able to show that the entropy of the black hole 
is the same as the statistical entropy of the effective string,
written in term of the levels of the effective string states.
In the BPS limit, the entropy of our black hole is finite.
We emphasize that for arbitrary compactification radii,
the entropy of the black hole 
which is constructed by 
two intersecting D-branes with no momentum 
vanishes in the BPS limit.

Further, we have studied the graybody factor with two large charges.
The graybody factor is proportional to the entropy 
of the black hole.
We have shown that the entropies of the black holes 
constructed by the intersecting D-branes are invariant 
under T-duality transformation~\cite{KS}.
Therefore, the graybody factors are also shown 
to be invariant 
under the T-duality transformation.

We have proposed that the compactification radii of the black holes 
are identified with the inverse of the surface gravities.
Using this determination, we have correctly 
obtained the integer valued 
Euler numbers which are calculated 
by using the ten-dimensional Gauss-Bonnet action.
Moreover, we have obtained the entropy formula 
of the black hole with no momentum.
In this paper, we have further obtained 
the interpretaion of the entropy 
of the black hole as the statistical entropy 
written in term of the level of states.

In the BPS limit, the entropies, the temperatures 
of the black holes 
are finite with two charges in our results.
The temperatures of the black hole with four charges 
vanish in the BPS limit with charges fixed.
However, the cross-sections and the entropies are infinite,
then it is rarely that the black holes become these states.
It is natural that they emit the charged particles 
and become the black holes with two charges.
Therefore, we think that 
the black holes with two charges are very special,
and we further need to study the behavior 
of the black holes with two charges.

\acknowledgements 
We thank Y. Kitazawa for discussions and 
for carefully reading the manuscript 
and suggesting various improvements.



\end{document}